\documentclass[Preprint, 12pt]{elsarticle}
\usepackage{graphics}
\usepackage{graphicx}
\usepackage{epsfig}
\usepackage{subfigure}
\usepackage{amssymb}
\journal{Chemical Physics Letters}
\begin{document}
\begin{frontmatter}
\title{{\it Ab-initio} study of structural and vibrational properties of KN$_3$ under pressure}
\author{K. Ramesh Babu and G. Vaitheeswaran}
\address{Advanced Centre of Research in High Energy Materials (ACRHEM),
University of Hyderabad, Prof. C. R. Rao Road, Gachibowli, Andhra Pradesh, Hyderabad - 500 046, India\\
\vspace{0.6in}
*Corresponding author E-mail address: gvsp@uohyd.ernet.in\\
\hspace {1.2in} Tel No.: +91-40-23138709\\
\hspace {1.2in} Fax No.: +91-40-23010227}
\end{frontmatter}
\clearpage
\section*{Abstract}
Structural stability and vibrational properties of KN$_3$ have been studied under pressure using the first principles calculations based on the density functional theory (DFT) as well as with semi empirical dispersion correction to the DFT to treat van der Waals interactions. The structural properties calculated through the dispersion correction are in good agreement with the experimental results when compared with the standard density functional theory. The lattice mode and the azide ion bending mode frequencies are found to be well reproduced by our calculations. As pressure increases, the E$_g$ mode gets softened as observed in the experiment. Despite the softening of E$_g$ mode, the enhancement of intramolecular interactions keeps the lattice stable.\\
Keywords:
equation of state; bulk modulus; vibrational frequencies
\newpage
\section{Introduction}
\label{}
Metal azides are well known energetic materials that find applications as primary explosives and gas generators \cite{Evans, Fair}. The structural simplicity of these materials enable them as model systems to study fast chemical reactions in solids with complex chemical bonding \cite{Bowden}. Among the metal azides, alkali metal azides are relatively stable compounds and received much interest towards their behavior at extreme conditions. Recent investigations on alkali metal azides revealed that they are potential candidates to generate green energy density material of a polymeric form of nitrogen \cite{Eremets-1, Eremets-2}. By using NaN$_3$ as a starting material Eremets et al., successfully synthesized polymeric nitrogen at pressures above 120 GPa \cite{Eremets-3}. High pressure studies on LiN$_3$ revealed that the system is stable up to the studied pressure range of 60 GPa and the formation of polymeric networks of nitrogen could be observed by compressing the material beyond 60 GPa \cite{Medvedev-1}. In view of this important application of these materials, it would be of interest to know the possibility of formation of single bonded network of nitrogen in other alkali azides. Hou et al., \cite{Hou} carried out X-ray diffraction measurements of cesium azide CsN$_3$ at high pressures and found that the tetragonal CsN$_3$ undergoes a series of structural phase transitions from ambient tetragonal $\rightarrow$ monoclinic (C2/m) at 0.5 GPa $\rightarrow$ monoclinic (P2/m) at 4.4 GPa $\rightarrow$ trigonal (P1) at 15.5 GPa respectively. The high pressure behavior of the other alkali metal azide KN$_3$, which is iso-structural with CsN$_3$, was experimentally studied through powder X-ray diffraction measurements \cite{Chen} at pressures up to 37 GPa. The authors found that the system may undergo a structural phase transition around 15.5 GPa but they could not identify the crystal structure of high pressure phase. Medvedev et al., \cite{Medvedev-2} performed a combined experimental study of powder X-ray diffraction and Raman spectroscopy and claimed that the system is structurally stable up to the pressure of 20 GPa despite the softening of the E$_g$ librational mode. In order to settle the controversy between the two above mentioned high pressure experiments there is a need to study the high pressure behavior of KN$_3$ system through accurate theoretical methods. Ab initio calculations based on density functional theory is a powerful tool in predicting the behavior of the solid systems at ambient as well as at high pressures. In continuation with our earlier work on monoclinic LiN$_3$ system\cite{Ramesh}, we further extend the study on high pressure behavior of tetragonal KN$_3$ system through first principles plane wave pseudo potential method including the van der Waals (vdW) interactions. The remainder of the paper is organized as follows: the next section deals with the computational details. Section 3 deals with the results and discussion and at the end we have given brief conclusions of our study.
\section{Computational details}
\label{}
All the calculations were performed within density functional theory using CASTEP code which uses a plane wave expansion in reciprocal space \cite{Payne, Segall}. The kinetic energy cut-off value for plane waves expansion was selected as 520 eV. The energy calculations in the first irreducible Brillouin zone were performed with a 5x8x5 k-point mesh using Monkhorst-Pack scheme \cite{Monkhorst}. Vanderbilt type ultrasoft pseudo potentials were employed to represent the Coulomb interaction between the valence electrons and ion core \cite{Vanderbilt}. The exchange-correlation energy was evaluated with local density approximation (LDA) of Ceperly-Alder \cite {Ceperley} and Perdew- Zunger \cite{Perdew-1} method and also with the generalized gradient approximation (GGA) of Perdew-Burke-Ernzerhof method \cite{Perdew}. Since KN$_3$ is a molecular crystal where van der Waal forces (vdW) play a major role in determining the structural parameters, the standard functional such as LDA, GGA would not sufficient to describe the system completely. Therefore in order to treat the vdW forces we have used semi-empirical dispersion correction scheme proposed by Grimme with PBE functional (PBE+G06) \cite{Grimme}. The calculations were performed by adopting the experimental crystal structure of KN$_3$ \cite{Muller}, a=6.113 $\AA$, c=7.094 $\AA$ with four molecules per unit cell (two units per primitive cell) as the initial structure, and it is relaxed to allow the ioinc configurations, cell shape, and volume to change at ambient pressure. Starting from the optimized crystal structure of KN$_3$ at 0 GPa, we applied hydrostatic pressure up to 20 GPa. Under a given pressure, the internal co-ordinates and unit cell parameters of the potassium azide were determined by minimizing the Hellmann-Feynmann force on the atoms and the stress on the unit cell simultaneously. In the geometry relaxation, the self-consistent convergence on the total energy is 5x10$^{-7}$eV/atom and the maximum force on the atom is found to be 10$^{-4}$eV/$\AA$. The vibrational frequencies have been calculated from the response to small atomic displacements within the linear response approach as implemented in CASTEP code \cite{Gonze, Refson}.
\section{Results and discussion}
\subsection{Structural properties}
The crystal structure of potassium azide is body-centered tetragonal with the symmetry of the space group D$^{18}_{4h}$-I4/mcm. As a first step, we performed full structural optimization of the unit cell and atomic positions within LDA (CA-PZ), GGA (PBE), GGA (PBE+G06). The optimized crystal structure of KN$_3$ within GGA (PBE+G06) is shown in Figure 1. The equilibrium lattice constants and volume of KN$_3$ are given in Table 1 together with experimental values. When compared to the experimental results, the general features of LDA, GGA are apparent. LDA over binding results in lattice parameters a and c that are 3.5$\%$ and 6.2 $\%$ smaller than experimental values while GGA relaxes the binding and improves the agreement with experiment by giving a and c 1.5$\%$ larger values than experimental ones. Since LDA and GGA does not treat the van der Waals forces in to account, the results could be improved by the inclusion of vdW forces in the calculations. In order to treat the vdW interactions, we have used the semi-empirical dispersion correction proposed by Grimme within PBE functional of GGA. The results obtained by PBE+G06 significantly improve the agreement with experiment when compared to the general LDA and GGA by reducing the error in lattice parameters to 0.1$\%$ smaller `a' and 1.8$\%$ smaller `c' with respect to the experimental values. It is observed that the calculated volume using PBE+G06 functional is closer to the experiment (smaller by 2.2$\%$) than LDA (smaller by 12$\%$) and GGA (larger by 4.6$\%$). Therefore for all further calculations we used both PBE+G06 functional in comparison with the PBE functional. The equation of state calculated from the hydrostatic compression simulation up to 20 GPa is compared with the experimental data of Cheng Ji et al.,\cite{Chen} and Medvedev et al.,\cite{Medvedev-2} in Figure 2 (a). Our calculations could qualitatively reproduce the trend of pressure-volume relation reported by the recent experiments \cite{Chen, Medvedev-2}. Especially the PBE+G06 results are in excellent agreement with the experimental data of Cheng Ji et al., at low pressures. But at high pressures GGA (PBE) calculations show very good agreement with data of Cheng Ji et al., compared to PBE+G06 values. This implies that at high pressures vdW forces does not play much role in KN$_3$ system. The bulk modulus (B$_0$) value was calculated to be 20.4 GPa using PBE and 26.3 GPa with PBE+G06 functional. Both the B$_0$ values are found to be in excellent agreement with that of experimental values 18.6 GPa \cite{Chen} and 27.4 GPa \cite{Medvedev-2}.
\paragraph*{}
The calculated lattice constants of KN$_3$ as functions of pressure are compared with experimental data in Figure 2 (b). The lattice constant `a' show better agreement with experiments when calculated with the PBE+G06 functional while PBE results show good agreement for `c'. This behavior can be understood from the fact that the percentage of error in the zero pressure value of `c' is larger for PBE+G06 functional compared to the PBE functional. In Figure 2 (c) we have shown the pressure dependence of a/a$_0$ and c/c$_0$ of KN$_3$ along with experimental data. Our calculations clearly show that the compressibility of KN$_3$ is anisotropic as the c-axis is more compressible than a-axis. This result is also consistent with both the experiments.
\subsection{Vibrational frequencies}
\subsubsection{Phonon frequencies at ambient pressure}
The primitive cell of KN$_3$ consists of eight atoms and hence it has 24 vibrational modes out of which three are acoustic modes and 21 are optical modes. According to the symmetry analysis of the point group D$_{4h}$, the acoustic and optical modes at the $\Gamma$ point can be classified into the following symmetry species:\\
$\Gamma_{aco}$ =  A$_{2u}$ + 2E$_{u}$\\
$\Gamma_{opt}$ = 2A$_{2g}$ + 2A$_{2u}$ + A$_{1g}$ + 4E$_{g}$ + 8E$_{u}$ + B$_{1g}$ + 2B$_{1u}$ + B$_{2g}$\\
 In this the A$_{2g}$ vibrations are silent as they do not cause a change in polarizability or dipole moment and therefore these modes are optically inactive. The modes E$_{g}$, A$_{1g}$, B$_{2g}$, and B$_{1g}$ are Raman active (RA) whereas the modes A$_{2u}$, B$_{1u}$, and E$_{1u}$ are infrared active (IA). The calculated vibrational frequencies at ambient pressure using PBE and PBE+G06 functionals are presented in Table 2 along with the experimental data. Since the KN$_3$ crystal consists of tightly bound azide ions which are loosely bound to K atoms, the vibrations involving the nitrogen atoms of azide ion could be labeled as internal vibrations and the external or lattice vibrations are those in which the azide ion move as rigid units along with the K sub lattice. In the calculated frequencies of KN$_3$, the modes from 89 cm$^{-1}$ to 200 cm$^{-1}$ are lattice modes for which the the results of PBE+G06 functional are in fair agreement with experiment \cite{Hathaway, Massa, Bryant} than PBE values. This is because the intermolecular van der Waals forces strongly couples in lattice modes which PBE could not take into account. But for the bending mode frequencies of internal modes due to the azide ions we find a very good agreement with experiment \cite{Lam, Iqbal, Papazian} using both PBE and PBE+G06 functionals. This could be because of the fact that vdW forces do not play much role in these vibrations as they are purely due to the motion of individual N atoms of each azide ion. However the stretching mode frequencies of azide ion ranging from 1200 cm$^{-1}$ to 1900 cm$^{-1}$ are underestimated by 8.4$\%$ for A$_{1g}$ and B$_{1g}$ modes (both are due to the symmetric stretching of azide ion) and the E$_{u}$ mode (asymmetric stretching of azide ion) is underestimated by 8.4$\%$. This might be due to the linear response approach which is solely based on the harmonic approximation and therefore the anharmonicity present in the higher frequencies which are mainly due to the azide ion could not be dealt by the present calculations.
 \subsubsection{Effect of pressure on phonon frequencies}
In this section, we discuss about the pressure induced phonon frequency shift under the application of hydro static compression. In Figure 3 (a) we have shown the pressure dependence of lattice modes  E$_{g}$(T), E$_{g}$(R) and  B$_{1g}$ (R). The frequencies of all the external modes increases with pressure except librational E$_{g}$(T) mode. The E$_{g}$ phonon mode, which is due to the translation of the K metal ion in the plane perpendicular to the z-axis as shown in Figure 4, softens under pressure with a pressure coefficient of -0.26 with PBE and -1.01 with PBE+G06, as observed in experiment \cite{Medvedev-2}. All the other external modes hardens as a function of pressure as shown in Figure 3 (b) and the corresponding pressure coefficients are presented in the Table 2. The calculated pressure coefficients for B$_{1g}$ (R) and E$_{1g}$ (R) are in close agreement with the experimentally reported values of Medvedev et al \cite{Medvedev-2}. The pressure dependence of bending modes of azide ion is shown in Fig 5 (a). We find an interesting feature of decrease in frequency of the A$_{2u}$ symmetric bending mode with pressure and the normal mode representation of this mode is shown in Figure 4. This kind of decrease in symmetric bending mode frequency of the azide ion with pressure was earlier observed in the case of AgN$_3$ \cite{Heming} and LiN$_3$ below a pressure of 30 GPa \cite{Ramesh}. So the decrease in  A$_{2u}$ mode frequency is not surprising and it merely suggests that the pressure enhances the interaction between the N atoms in the azide ion. In Figure 5 (b) and 5 (c) we have shown the Raman active symmetric (B$_{2g}$, A$_{1g}$) and Infrared active asymmetric (E$_{u}$) stretching mode behavior of the azide ion under pressure. The frequency of these three stretching modes increases monotonically with pressure with a pressure coefficients as shown in Table 2. As observed in experiment \cite{Medvedev-2} we could not find any discontinuity in these internal Raman active symmetric stretching modes under pressure. Overall from the study of vibrational frequencies of optical modes of KN$_3$, it is clear that the pressure coefficients of external modes are higher than that of internal modes implying the enhancement of intermolecular interactions under pressure. It is also clear that despite the softening of  E$_{g}$(T) mode, the lattice is stable due to the strong increase in intramolecular interactions under the studied pressure range. From our study we propose that for further compression the tetragonal KN$_3$  crystal may show lattice instability.
\paragraph*{}
In order to understand the volume dependence of the vibrational modes we have calculated the Gr\"{u}neisen parameter ($\gamma_i$) of each mode. The Gr\"{u}neisen parameter can be calculated by using
\begin{equation}
\gamma_i = \frac{\partial(ln\nu_i)}{\partial(lnV)}
\end{equation}
The calculated $\gamma_i$ for each mode is presented in Table 3. The $\gamma_i$ for E$_{g}$(T) and A$_{2u}$ mode are obtained to be negative, whereas the value is positive for all the other modes. It can also be noticed that the magnitude of $\gamma_i$  is larger for lattice modes than those of internal modes involving azide ion vibrations alone.
\section{Conclusions}
We have studied the structural stability and vibrational properties of potassium azide KN$_3$ under high pressure up to 20 GPa using first principles theory including van der Waals interactions. The calculated ground state properties are in good agreement with experiment using PBE+G06 functional when compared with PBE functional. The calculated bulk modulus was found to be in excellent agreement with experiment. The vibrational frequencies were calculated and are in fair agreement with experiment using PBE+G06 functional. At high pressures, the translational lattice mode E$_g$ (T) as well as the symmetric bending mode A$_{2u}$ gets softened. This softening of E$_g$ (T) mode suggests that the lattice may become unstable at high pressures but due to the increase in intra molecular interactions between the N atoms, we could not observe any lattice instability up to the studied pressure range.

\newpage
\begin{figure}
\centering
\includegraphics[width=12cm]{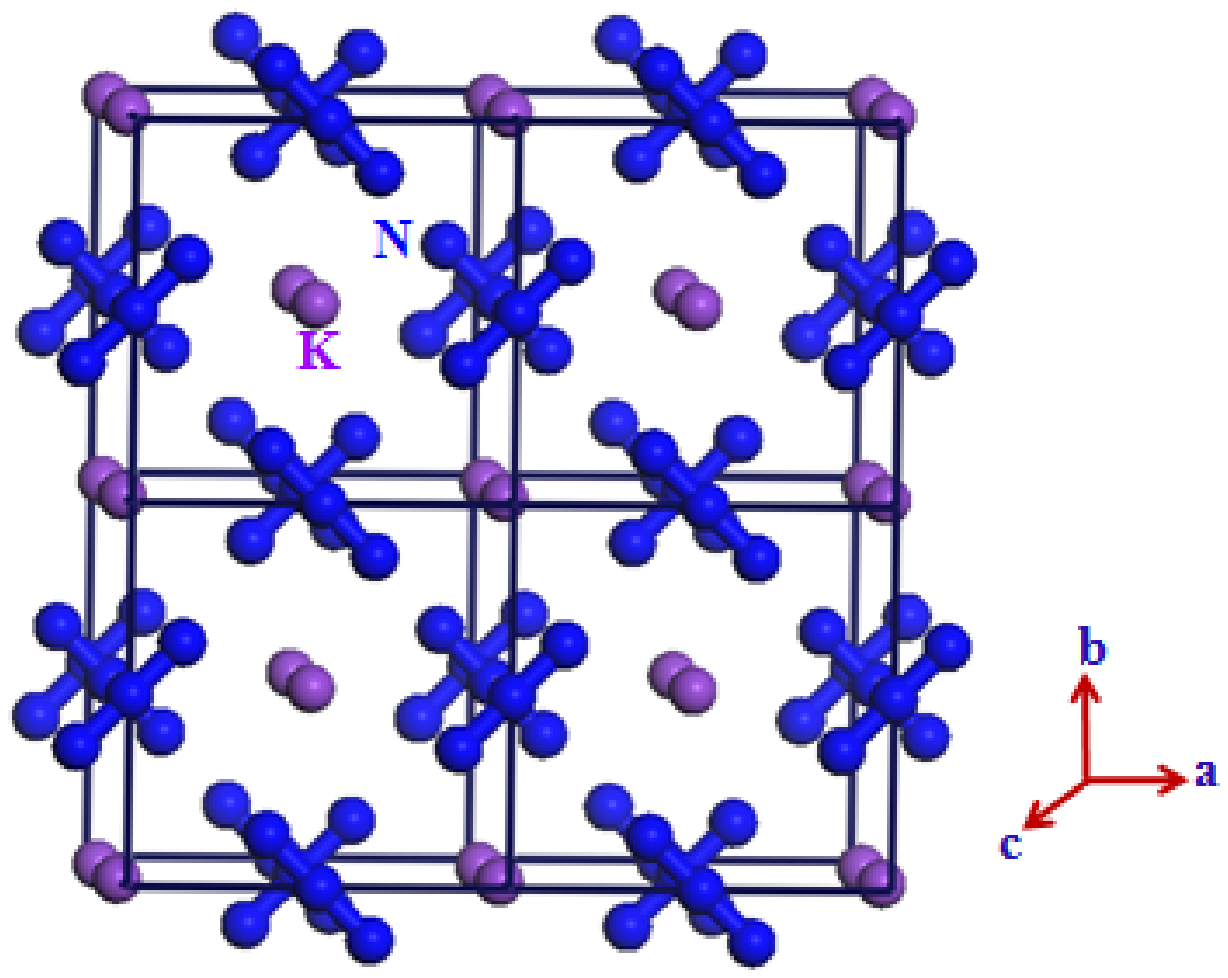}\\
\caption{(Colour online) Optimized crystal structure of KN$_3$ within PBE+G06 functional.} \label{Fig2}
\end{figure}
\clearpage
\newpage
\begin{figure}[h!]
\begin{center}
\subfigure[]{
\includegraphics[scale=0.25]{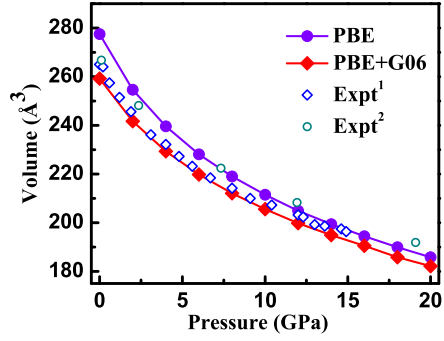}}
\subfigure[]{
\includegraphics[scale=0.25]{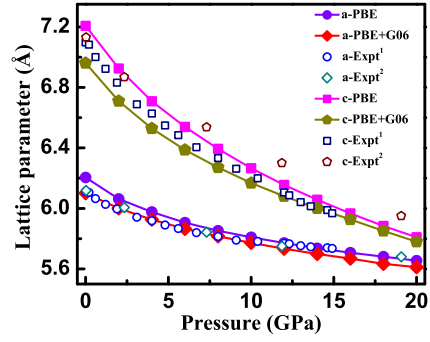}}
\subfigure[]{
\includegraphics[scale=0.25]{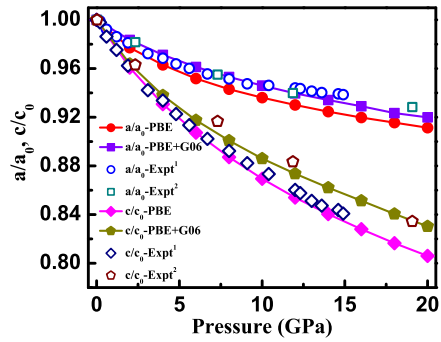}}
\caption{(Colour online) (a) Equation of state of KN$_3$. Experimental values are taken from $^1$Ref \cite{Chen} and $^2$Ref\cite{Medvedev-2} (b) Lattice parameters variation of KN$_3$ with pressure along with experimental data taken from $^1$Ref\cite{Chen} and $^2$Ref\cite{Medvedev-2} (c) Normalized lattice parameters variation of KN$_3$ with pressure. Experimental values are taken from $^1$Ref \cite{Chen} and $^2$Ref\cite{Medvedev-2} }\label{Fig 1}
\end{center}
\end{figure}
\clearpage
\newpage
\begin{figure}[h!]
\begin{center}
\subfigure[]{
\includegraphics[scale=0.4]{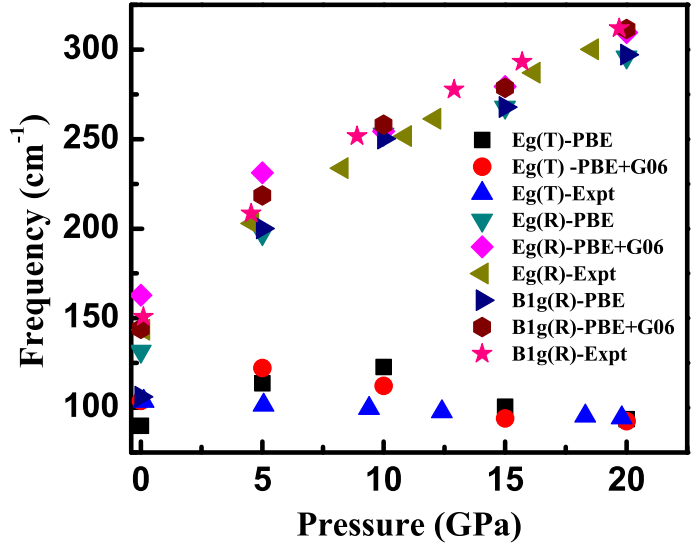}}
\subfigure[]{
\includegraphics[scale=0.4]{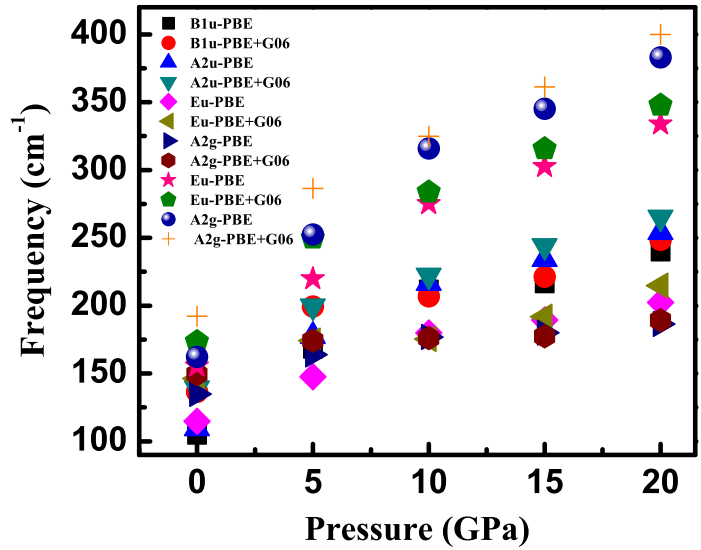}}
\caption{(Colour online) (a) The dependence of E$_g$(T), E$_g$(R), and B$_{1g}$(R) mode frequencies of KN$_3$ with pressure. Also shown experimental data of Medvedev et al., Ref\cite{Medvedev-2} (b) Effect of pressure on external or lattice mode frequencies of KN$_3$.}
\end{center}
\end{figure}
\clearpage
\newpage
\begin{figure}
\centering
\includegraphics[width=12cm]{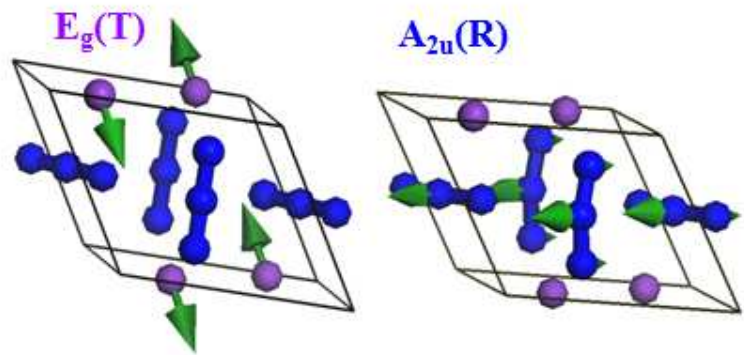}\\
\caption{(Colour online) Normal mode representation of two softening modes E$_g$ (T) and A$_{2u}$ of KN$_3$.}
\end{figure}
\clearpage
\newpage
\begin{figure}[h!]
\begin{center}
\subfigure[]{
\includegraphics[scale=0.32]{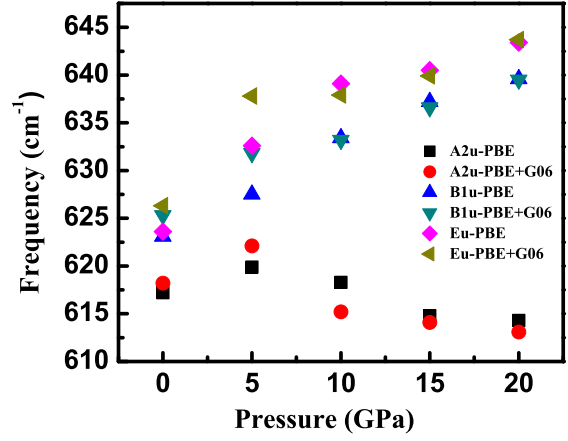}}
\subfigure[]{
\includegraphics[scale=0.24]{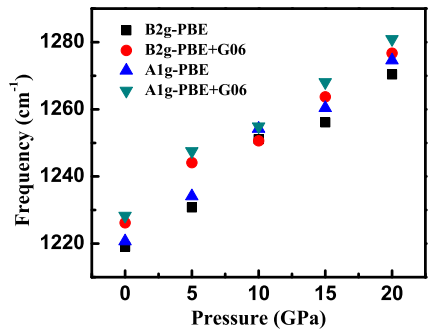}}
\subfigure[]{
\includegraphics[scale=0.24]{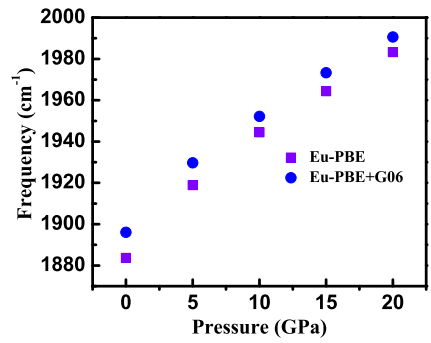}}
\caption{(Colour online) (a) Variation of bending mode frequencies of KN$_3$ with pressure (b) and (c) Pressure effect on stretching mode frequencies of KN$_3$.}
\end{center}
\end{figure}
\clearpage
\newpage
\begin{table}
\caption{The calculated ground state properties of tetragonal KN$_3$. Experimental data are taken from Ref \cite{Muller}}
\label{tab.1}
\begin{center}
\begin{tabular}{cccccc}\hline \hline
 Parameter &\hspace{2.0cm}Theory&  & &Expt\\
 & CA-PZ & PBE & PBE+G06 &\\ \hline
 a (\AA) &5.898& 6.205&6.102&6.113\\
 c (\AA) &6.653 & 7.207  &6.961 &7.094 \\
 V (\AA$^3$) & 231.38 & 277.54  &259.16&265.09 \\
 K  & (0 0 0.25) & (0 0 0.25) &(0 0 0.25)&  (0 0 0.25)\\
 mid N  &(0 0.5 0) & (0 0.5 0)&(0 0.5 0)& (0 0.5 0)\\
 end N &(0.1450 0.6450 0) &(0.1388 0.6388 0)&(0.1411 0.6411 0)& (0.1358 0.6358 0)\\
 B$_0$&-- &20.4 & 26.3&18.6$^a$,27.4$^b$\\
 B'$_0$&-- &3.96 & 3.93&6.7$^a$,4.4$^b$\\ \hline\hline
\end{tabular}
$^a$\cite{Chen}, $^b$\cite{Medvedev-2}
\end{center}
\end{table}
\clearpage
\newpage
\begin{table}
\caption{The calculated vibrational frequencies (cm$^{-1}$) of tetragonal KN$_3$ at ambient pressure and their corresponding pressure coefficients, P$_c$ in (cm$^{-1}$/GPa) within PBE and PBE+G06 functionals. T and R corresponds to Translational and Rotational vibrations respectively. IA and RA infers the Infrared active and Raman active modes of KN$_3$.}
\label{tab.2}
\begin{center}
\begin{tabular} { l | c  c | c |c r }\hline
Mode symmetry& &Frequency (cm$^{-1}$)& Expt&P$_c$  & \\\hline
& PBE & PBE+G06& &PBE & PBE+G06  \\
E$_g$(T)  (RA)&89.8&103.8&103$^1$&-0.26&-1.01\\
B$_{1u}(T)$&104.9&136.5& &6.36&4.92\\
B$_{1g}$(R)  (RA)&106.1&139.2& &8.98&7.89\\
A$_{2u}$(T)  (IA)&109.6&144.1&138.2$^2$ & 5.91&6.91\\
E$_{u}$(T)  (IA)&114.7&146.2& &4.35&3.11\\
E$_{g}$(R)  (RA)&131.5&149.4&145$^1$, 147$^3$ &7.99&6.88\\
A$_{2g}$(T)&134.9&162.8& &2.38& 1.66 \\
E$_{u}$(T)  (IA)&155.5&173.4&168$^2$ &8.78& 8.29\\
A$_{2g}$(R)&162.3&192.2& &10.67&9.80 \\
A$_{2u}$  (IA)&617.2&618.2&627$^4$, 643$^5$, 624$^6$&-0.21&-0.37 \\
B$_{1u}$&623.1&625.3& & 0.85& 0.66\\
E$_{u}$&623.6&626.3& 624$^4$, 647$^5$, 646$^6$&0.95& 0.74\\
B$_{2g}$  (RA)&1219.1&1226.3&1339$^1$&2.56& 2.41\\
A$_{1g}$  (RA)&1220.7&1228.5& 1340$^1$& 2.68& 2.52\\
E$_{u}$ (IA)&1883.7&1896.1&2002.2$^6$ &4.89& 4.24\\ \hline
\end{tabular}
$^1$\cite{Hathaway}, $^2$\cite{Massa}, $^3$\cite{Bryant}, $^4$\cite{Lam}, $^5$\cite{Iqbal}, $^6$\cite{Papazian}
\end{center}
\end{table}
\begin{table}
\caption{The calculated Gr\"{u}neisen parameters ($\gamma$) of the vibrational frequencies of all vibrational modes of tetragonal KN$_3$ within PBE and PBE+G06 functionals.}
\label{tab.1}
\begin{center}
\begin{tabular}{cccccc}\hline \hline
 Mode Symmetry & Gr\"{u}neisen parameter ($\gamma$) & & \\
               &   PBE   & PBE+G06\\ \hline
 E$_g$(T) (RA)& -0.13&-0.37\\
 B$_{1u}(T)$   &2.05&1.38\\
 B$_{1g}$(R) (RA)&2.54&1.88\\
 A$_{2u}$(T) (IA)&2.09&1.57\\
 E$_{u}$(T) (IA)&1.45&0.86\\
 E$_{g}$(R) (RA)&2.03& 1.54\\
 A$_{2g}$(T) & 0.79&0.52\\
 E$_{u}$(T) (IA) & 1.92&1.70\\
 A$_{2g}$(R) &2.13&1.78\\
 A$_{2u}$ (IA)& -0.01&-0.02\\
 B$_{1u}$ & 0.06&0.05\\
 E$_{u}$&0.07&0.06\\
 B$_{2g}$ (RA)&0.10&0.09\\
 A$_{1g}$ (RA)&0.11&0.09\\
 E$_{u}$ (IA)&0.13&0.12\\ \hline
\end{tabular}
\end{center}
\end{table}

\end{document}